\documentclass[%
 amsmath,amssymb,
 aip,
preprint,
]{revtex4-1}

\usepackage{listings}       
\usepackage[colorlinks=true,linkcolor=blue]{hyperref} 
\usepackage{graphicx} 
\usepackage{epstopdf}       
\usepackage[utf8]{inputenc} 
\usepackage[T1]{fontenc}
\usepackage{bm} 
\usepackage{mathptmx}
\usepackage{multirow}    

\newcommand{\relacsfdist}{\delta}

\begin{document}

\title[]{A Bin and Hash Method for Analyzing Reference Data and Descriptors in Machine Learning Potentials}
\author{Mart\'in Leandro Paleico}
\email{martin.paleico@uni-goettingen.de}
\author{J\"org Behler}
\email{joerg.behler@uni-goettingen.de}
\affiliation{Universit\"{a}t G\"{o}ttingen, Institut f\"{u}r Physikalische Chemie, Theoretische Chemie, Tammannstra\ss{}e 6, 37077 G\"{o}ttingen, Germany}
\date{\today}

\begin{abstract}
In recent years the development of machine learning (ML) potentials (MLP) has become a very active field of research. Numerous approaches have been proposed, which allow to perform extended simulations of large systems at a small fraction of the computational costs of electronic structure calculations. The key to the success of modern ML potentials is the close-to first principles quality description of the atomic interactions. This accuracy is reached by using very flexible functional forms in combination with high-level reference data from electronic structure calculations. These data sets can include up to hundreds of thousands of structures covering millions of atomic environments to ensure that all relevant features of the potential energy surface are well represented. The handling of such large data sets is nowadays becoming one of the main challenges in the construction of ML potentials.
In this paper we present a method, the bin-and-hash (BAH) algorithm, to overcome this problem by enabling the efficient identification and comparison of large numbers of multidimensional vectors. Such vectors emerge in multiple contexts in the construction of ML potentials. Examples are the comparison of local atomic environments to identify and avoid unnecessary redundant information in the reference data sets that is costly in terms of both the electronic structure calculations as well as the training process, the assessment of the quality of the descriptors used as structural fingerprints in many types of ML potentials, and the detection of possibly unreliable data points.
The BAH algorithm is illustrated for the example of high-dimensional neural network potentials using atom-centered symmetry functions for the geometrical description of the atomic environments, but the method is general and can be combined with any current type of ML potential.
\end{abstract}

\maketitle
\bigskip


\section{Introduction} \label{sec:introduction}

Machine-learning (ML) has become an important tool for the development of atomistic potentials, with a wide variety of applications in chemistry, physics and materials science.\cite{behler_perspective_2016,P4938,P5673}.  
Machine learning potentials, like many other applications of machine learning algorithms, aim at approximating unknown functions, which in the present case is the multidimensional potential energy surface (PES) of the system of interest as a function of the atomic positions. The required information is obtained from sampling the PES at discrete points, i.e. particular atomic configurations, utilizing comparably demanding electronic structure methods such as density functional theory (DFT)~\cite{hohenberg_inhomogeneous_1964, kohn_self-consistent_1965}. Once constructed, the ML potential can then be used to perform cheap simulations with first principles accuracy for systems of significantly increased size and for extended time scales, to address problems which are inaccessible, e.g., to ab initio molecular dynamics simulations.

Many types of ML potentials have been developed in recent years, including different flavors of artificial neural-network based potentials~\cite{blank_neural_1995,behler_generalized_2007,jiang_permutation_2013,P0421,P2059,P5577,P5366,P5596,P4945}, Gaussian approximation potentials~\cite{bartok_gaussian_2010,P4429}, moment tensor potentials~\cite{shapeev_moment_2016}, spectral neighbor analysis potentials~\cite{thompson_spectral_2015}, and many others \cite{P5420,P4479}. Apart from reproducing atomic interactions, machine learning methods have also seen increasing applications that attempt to predict derived properties instead of those directly associated with the PES, such as dipole moments~\cite{P5205,P2041,pereira_machine_2018}, charges~\cite{artrith_high-dimensional_2011,morawietz_neural_2012,P4945,P5313,P5885}, electronegativities~\cite{P4990}, band gaps~\cite{lee_prediction_2016, pilania_multi-fidelity_2017}, spins~\cite{P5867}, and atomization energies~\cite{rupp_fast_2012}.
All these applications of ML algorithms rely on the availability of large reference data sets that are used to train the respective ML method to reliably reproduce the property of interest. Generating these data sets is computationally very demanding, and thus the amount of data should be kept as small as possible, which is a very challenging task. In the present work we address this by introducing the bin and hash (BAH) algorithm, enabling a computationally very efficient analysis of large data sets. This analysis is possible before training of the ML algorithm of choice has been performed, and even before the electronic structure calculations are carried out, which allows to guide the selection of the most important structures. 

Data set maintenance and analysis as well as atomic fingerprint selection, i.e. finding suitable representations of atomic geometric environments, have been active areas of research accompanying the rise in popularity of ML methods. The use of large and increasingly automatically generated data sets and algorithms to programatically explore PESs~\cite{partay_efficient_2010, kolsbjerg_neural-network-enhanced_2018, jennings_genetic_2019} has led to the need for tools that can deal with the amount and complexity of data. One such method is the dimensionality reduction algorithm SketchMap~\cite{ceriotti_simplifying_2011, de_mapping_2017}, which can be utilized to group structures together into similarity clusters. More direct tools measuring distances in configuration space~\cite{sadeghi_metrics_2013} and structural similarities of solids~\cite{zhu_fingerprint_2016} are also useful for analyzing collections of structures. Previous attempts based on ML descriptors such as SOAPs~\cite{de_comparing_2016} have also been successful at establishing a similarity measurement algorithm, and recently a more generalized study has been published, looking at the most common ML descriptors~\cite{parsaeifard_assessment_2020} and their relative behavior in describing atomic environments as well as the relationships between property space (in this case energy) and distances in descriptor space.

As an inherent part of most MLP approaches, atomic fingerprint selection, has also attracted a lot of attention. In the wider field of machine learning this is done with meta-analysis methods, such as hyperparameter optimization~\cite{hutter_beyond_2015, luo_review_2016, klein_fast_2017}. Unfortunately these methods are usually rather complex and expensive, requiring multiple training and fitting iterations, which precludes their use for large MLP data sets. Methods specifically designed for MLP also exist, that attempt to refine the contents of these atomic fingerprints. Among them we find attempts at utilizing genetic algorithm optimization~\cite{P5292,browning_genetic_2017} to select the best fingerprint sets through evolution, or CUR decomposition~\cite{imbalzano_automatic_2018} to select fingerprints through dimensionality reduction.

In this work we use high-dimensional neural network potentials (HDNNP) as proposed by Behler and Parrinello in 2007~\cite{behler_generalized_2007, behler_first_2017} to illustrate our algorithm, but the algorithm is very general and can be used in combination with many other types of ML potentials and atomic environment descriptors. The main idea of the HDNNP approach, which is also used in most other classes of high-dimensional ML potentials, is the construction of the total potential energy $E$ of the system as a sum of atomic energy contributions $E_i$ from all $N_{\mathrm{atom}}$ atoms in the system as
\begin{eqnarray}
E=\sum_{i=1}^{N_{\mathrm{atoms}}} E_i \quad .
\end{eqnarray}
These atomic energies depend on the local chemical environments up to a cutoff radius $R_{\mathrm{c}}$, which has to be chosen large enough to capture all energetically relevant atomic interactions. Typically cutoff values of 6-10~\AA{} are used. The positions of all neighboring atoms in the resulting cutoff sphere must be provided to individual element-dependent atomic neural networks yielding the atomic energies. Many types of descriptors are available in the literature \cite{bartok_representing_2013,P5356,P5075,P5420,P5292,P5645,P5318}, and the most frequently used type in the context of HDNNPs are atom-centered symmetry functions (ACSFs)~\cite{behler_atom-centered_2011}, which form a vector $\mathbf{G}_i$ of input coordinates for each atomic neural network that is invariant with respect to rotation, translation and permutation, i.e. the order of the atoms in the system. A detailed discussion of the functional forms of ACSFs and their properties can be found in Ref.~\citenum{behler_atom-centered_2011}, and here we just use them as placeholders for any ordered set of descriptor values that provides a meaningful structural fingerprint of the local atomic environments. 

The atomic neural networks represent the analytic functional form of the HDNNP and contain a large number of fitting parameters, the neural network weights, which are optimized in an iterative training process to reproduce a given reference data set of energies and forces for representative systems obtained from electronic structure calculations. Once the HDNNP has been trained using this data, the energies and forces of a large number of configurations can be computed at a small fraction of the computational costs of the underlying electronic structure method, which enables extended molecular dynamics and Monte Carlo simulations of large systems with close-to first-principles quality. For all details about the method, the training process and the validation strategies for HDNNPs we refer the interested reader to a series of recent reviews~\cite{P4106,behler_constructing_2015,behler_first_2017}.

The construction of HDNNPs involves the use of large amounts of data, and the generation of the reference electronic structure data often represents the computationally most demanding step. It is therefore desirable to reduce the amount of data as much as possible by only including those structures -- or more specifically atomic environments -- which are different enough from the data already included in the reference set to justify the effort of an electronic structure calculation. In addition, also the training process of the HDNNP becomes more time consuming with increasing amount of data. In recent years, active learning\cite{P5900} has become a standard procedure to identify the most relevant structures \cite{artrith_high-dimensional_2012,P4939,P5782,P5842}. Still, the inclusion of a wide range of structurally different atomic environments in the training process is essential for the construction of a reliable HDNNP, as the underlying functional form is non-physical, and the correct physical shape of the potenial-energy surface can only be learned if all of its relevant features are included in the training set. Consequently, for each system a compromise between the effort of constructing large data sets and the accuracy and range of applicability of the HDNNP has to be found.

The use of large amounts of data poses several challenges. First, a set of ACSF descriptors has to be defined for each element in the system to construct structural fingerprints that can be used by the atomic neural networks to construct the energy expression of the HDNNP. These ACSFs can be used for the quantification of the similarity of different atomic environments. Typically, a set of 20-100 ACSFs is used for this purpose, which depend on parameters defining their spatial shape~\cite{behler_atom-centered_2011}. Second, to keep the data sets small, the inclusion of redundant information has to be avoided, which requires an efficient analysis and comparison of the local chemical environments of the atoms given by the ACSF vectors. As we will see below, naive pairwise comparisons are not a viable option for the typical data sets consisting of tens of thousands of structures, each containing up to a few hundred atoms. Third, the costs of the reference electronic structure calculations should be kept as low as possible, but numerical noise that can arise, e.g., from loose but time-saving settings of the electronic structure codes must be avoided. Substantial noise in the data represents contradictory information, which prevents a smooth convergence of the fitting process to low root-mean squared errors for the energies and forces. 

In this paper, we propose a simple, fast and efficient algorithm based on the well known hash table~\cite{cormen_introduction_2009} data structure. The algorithm is described in Sec.~\ref{sec:compmet_bah}. We use the vector of ACSF values belonging to an atomic environment, the same vector that an atomic neural network in HDNNPs would receive as an input, but we first pre-process it by a bin and hash approach. Binning is described in Sec.~\ref{sec:compmet_bah_binning}, and the procedure of hashing and the workings of hash tables in Sec.~\ref{sec:compmet_bah_hashing}. This creates a numerically unique representation of each environment, where searches for repeated representations are fast and scale well with the number of environments under consideration. In addition, this procedure does not depend on the availability of a trained HDNNP, which is an advantage compared to active learning strategies. The procedure is very fast, and we benchmark it in relation to a naive direct comparison approach in Sec.~\ref{sec:compmet_bah_naive}, with big $O$ notation~\cite{cormen_introduction_2009} scaling discussed in Sec.~\ref{sec:compmet_bah_scaling}.

In Sec.~\ref{sec:results_bah}, we show results from the application of the algorithm. Concrete timings are presented in Sec.~\ref{sec:results_bah_timings}, confirming the scaling expected based on theoretical considerations. Section~\ref{sec:results_bah_distance} demonstrates how the BAH algorithm reproduces distances in ACSF vector space, while Sec.~\ref{sec:results_bah_sfsets} shows the behavior of the algorithm when changing the number of binning subdivisions and the ACSF set description of the data set, and how this can be utilized to qualitatively evaluate the suitability of a given ACSF set, without requiring the lengthy process of previously fitting a potential. Finally, Sec.~\ref{sec:results_bah_contradictions} shows how the method can be easily utilized to find similar atomic environments and contradicting information in a data set. 

Overall these applications are examples for the well known and complex problem of efficiently finding distances and nearest neighbors in points belonging to multi-dimensional data. Previous approaches include making use of complex binary tree data structures such as kDtrees~\cite{bentley_multidimensional_1975, cormen_introduction_2009}, that can efficiently store data points according to their mutual distance in multi-dimensional space and rapidly reduce a search space due to their binary structure; and dimensionality reduction algorithms such as principal component analysis (PCA)~\cite{pearson_lines_1901, hotelling_analysis_1933} and SketchMap~\cite{ceriotti_simplifying_2011} that instead reduce the size of the space under consideration. All of these algorithms are very powerful and suited for their particular applications, but are often too complex and slow for the current goal. Our BAH approach is fast and simple, and works in principle for any dimensionality. It simplifies the process of dimensionality reduction by performing a reduction evenly across the coordinate space instead of centering on the most important directions like PCA and SketchMap.

\section{The Bin and Hash Algorithm} \label{sec:compmet_bah}
 
\subsection{Description of the Algorithm} \label{sec:compmet_bah_description}

\begin{figure}
    \centering
    \includegraphics[width=\linewidth]{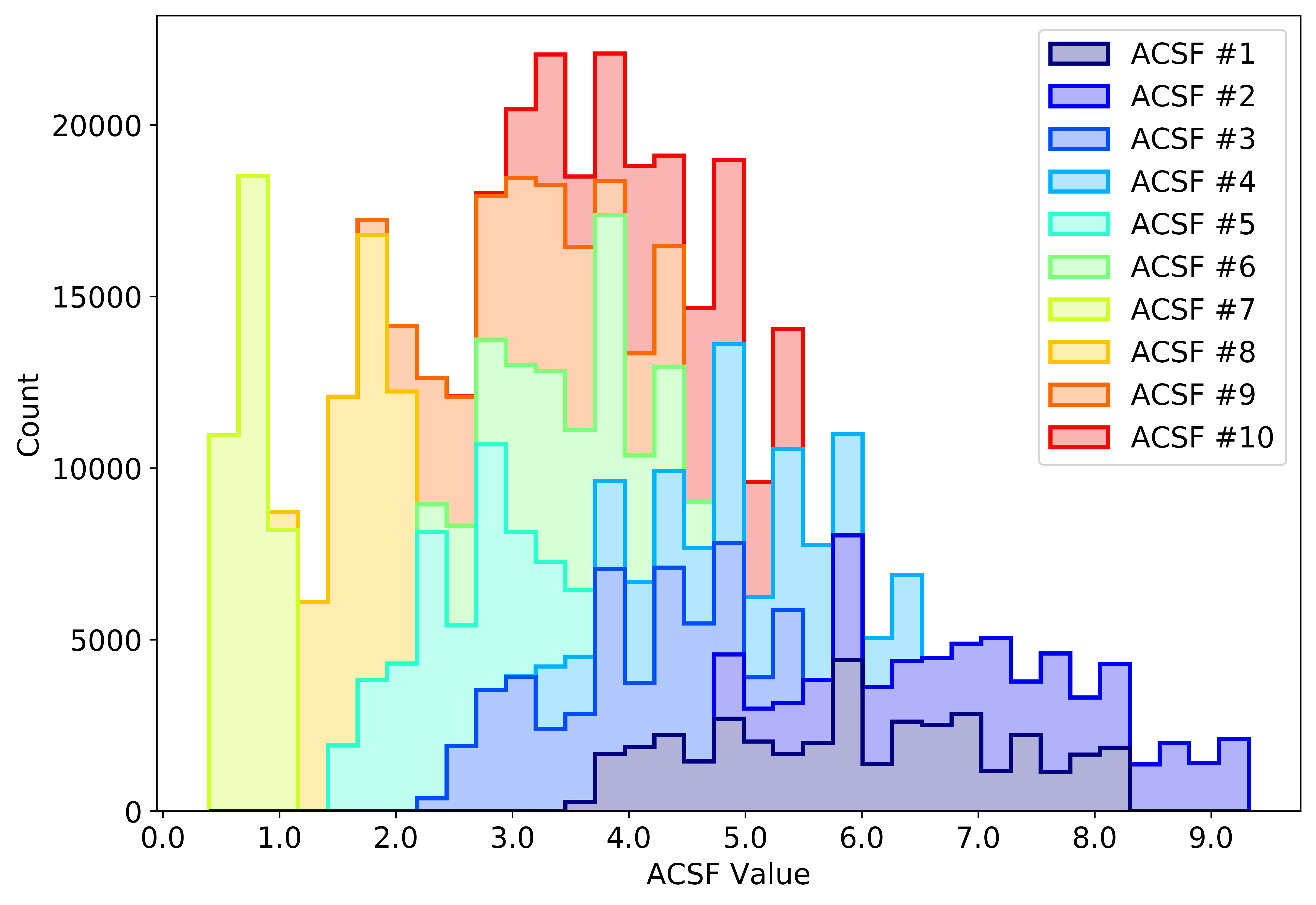}
    \caption{Stacked histogram plot of the values of the first 10 radial ACSFs in the ZnO data set describing the atomic environments of the oxygen atoms.}
    \label{fig:sf-histo}
\end{figure}


\lstset{emph={for, each, in, =, not, if, else},emphstyle=\textbf} 
\begin{figure} 
\begin{minipage}{\linewidth}
\renewcommand{\lstlistingname}{Code Block}
\begin{lstlisting}[caption={Pseudocode for the bin and hash algorithm},captionpos=b,label={code:bah-pseudocode},numbers=left,frame=single,basicstyle=\footnotesize]
divs =  number of subdivisions in ACSF space
for atom_env_i in data set
    for acsf_j in acsf_set
        calculate symmetry function vector Gi={Gj}
find Gjmax and Gjmin across each acsf component Gj
initialize empty hash table Ht
for each Gi vector
    bin Gi vector Bi={Bj}, 
        Bj=divs*(Gjmax-Gj)/(Gjmax-Gjmin)
    calculate hash Hi=hash(Bi)
    if Hi not in Ht
        store it Ht[Hi]=j index
    else
        count as collision ncolls+=1
        add to existing record in hash table
        Ht[Hi] append(j index)
\end{lstlisting}
\end{minipage}
\end{figure}

\begin{figure}
    \centering
    \includegraphics[width=\linewidth]{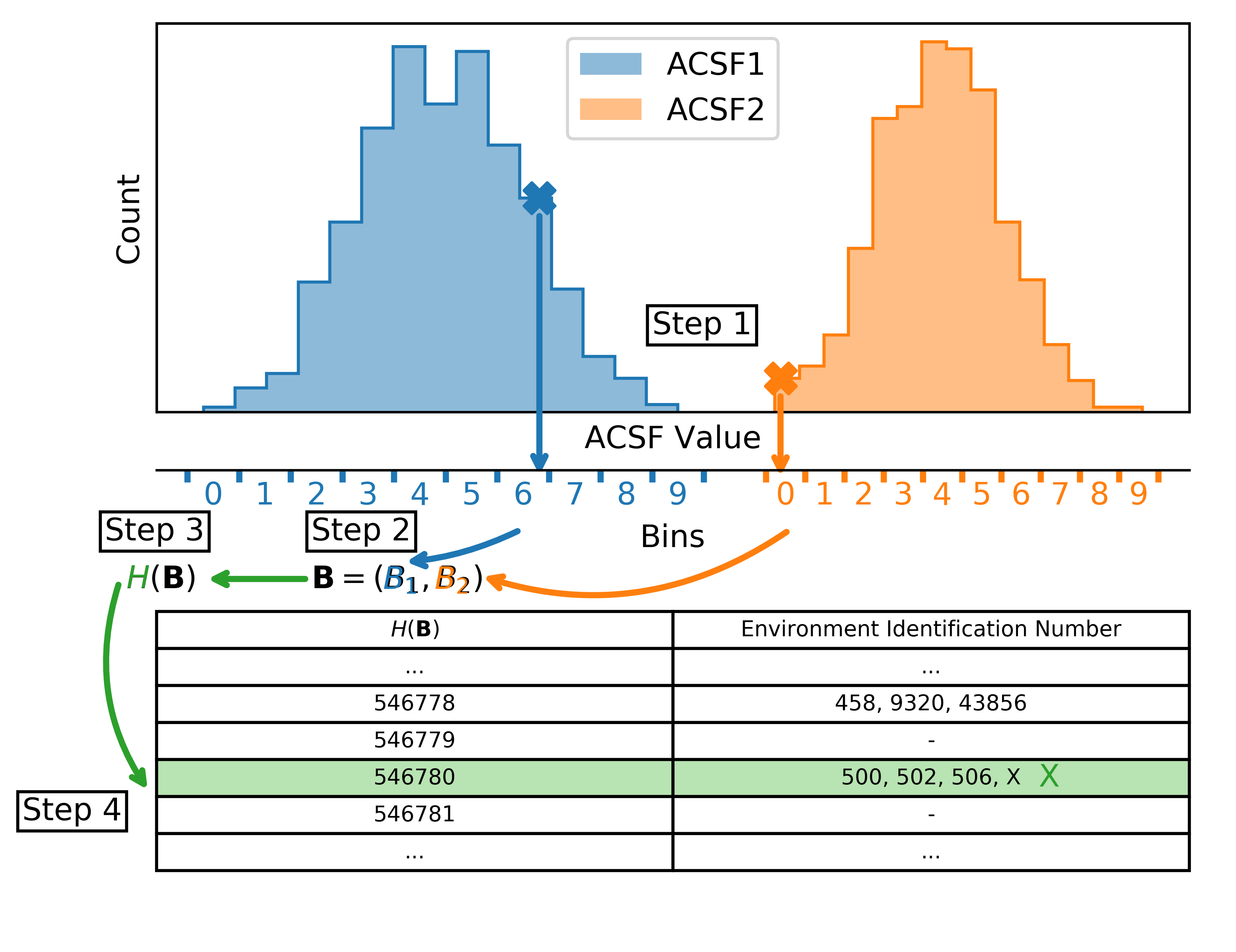}
    \caption{Illustration of the BAH approach: Each atomic environment in this example is characterized by a two-dimensional ACSF vector $\mathbf{G}=(G_1,G_2)$. In step 1, the histograms corresponding to the ACSFs are generated as a visualization aid. The values of $G_1$ and $G_2$ are highlighted by the crosses for one particular example environment. This ACSF vector is then binned to a pair of integer values, forming the binned vector $\mathbf{B}=(B_1,B_2)$ in step 2. In step 3 the hash $H(\mathbf{B})$ of this binned vector is calculated. Finally, in step 4 this hash is used (directly or indirectly) to index into the hash table, and add the atomic environment to a counter for similar environments.}
    \label{fig:bah-diagram}
\end{figure}

Here, we will first give a general overview about the bin and hash algorithm summarized as pseudocode in code block~\ref{code:bah-pseudocode}. The details of each of its components will be discussed in the following sections. 

As example system we choose zinc oxide. A typical distribution of ACSF values is presented in Fig.~\ref{fig:sf-histo} in the form of a stacked histogram plot, for the first 10 ACSFs of a small data set containing 1192 configurations of a ZnO($10\bar{1}0$) surface slab with in total 75360 atomic environments. The structures included in the data set consist of bulk cut slabs, relaxed slabs, and configurations extracted from MDs, with different number of layers. Overall, 58 distinct atom-centered symmetry functions are used per element to describe the atomic environments, and the parameters defining the ACSFs are given in the supporting information. We can see that even for such a relatively small data set the distribution of data already has a rather complex form.

The individual steps forming the BAH algorithm are illustrated in Fig.~\ref{fig:bah-diagram}.
Starting from the histogram of ACSF values shown schematically, in a first step the range of each ACSF is split into a predefined number of subdivisions, typically between $10^1$ and $10^7$ bins, taking into account the maximum and minimum values present in the data set. This transforms the ACSF vector $\mathbf{G}_i$ for a given atomic environment $i$ from a float-based continuous representation to an integer-valued binned vector $\mathbf{B}_i$ of the same dimensionality (step 2). This binned vector is then hashed generating the one-dimensional hash key $H_i$ (step 3), which is then used for constructing a hash table ($H_t$) (step 4). The binning achieves two goals at once: getting rid of the floating point representation, which does not allow for an accurate transformation to a hash, since the hash would be numerically very sensitive to the round-off errors of the floating point values, and binning similar ACSF vectors to the same $\mathbf{B}_i$ vector, finally yielding the same hash key. The step of hashing the integer vectors into hash buckets enables a fast and efficient storage and lookup for large data sets. Both parts of the algorithm -- binning and hashing -- are thus vital for its performance.

Any $\mathbf{G}_i$ vectors that result in a hash collision, i.e. they end up in the same hash table bucket, are deemed to be similar, and -- depending on the number of subdivisions -- usually exactly the same apart from floating point round-off errors (see Sec.~\ref{sec:results_bah_distance}). The algorithm keeps track of the total number of collisions recorded for a data set and the maximum number of collisions for all the buckets. Additionally, every time a collision is detected the ID of the colliding atomic environment is stored in the hash table in the corresponding bucket, which enables to retrieve the colliding environments afterwards for analysis.

An obvious problem of this algorithm is that environments might be very close to the border between two bins. Given two very similar environments, both could be assigned to different bins resulting in completely different hash values, although the atomic configurations are essentially identical. In this case, two environments that should lead to a collision, do not. A straightforward solution to this problem is to use the algorithm with multiple different divisions of the ACSF domain, and to compare the obtained binning. In this way it can be excluded that very similar environments are converted to different hash keys. Still, even when using multiple binnings, the algorithm remains computationally very efficient.

\subsection{Analysis of the Algorithm} \label{sec:compmet_bah_analysis}

Next, we analyze the scaling of the algorithm. This scaling is of particular relevance given the sheer size of the typical data sets used in the construction of ML potentials. Many other more sophisticated algorithms work perfectly well when tested on small example cases, but scale very inefficiently for realistic data sets containing tens or even hundreds of thousands of structures, each consisting of many atomic environments. Initially, we comment on the possibility of utilizing neighbor lists. Then, we describe the naive approach of a brute force comparison as a reference, before discussing the behavior of the binning and hashing operations. Finally, we derive the scaling in big $O$ notation~\cite{cormen_introduction_2009}.

\subsubsection{Cell-Based Neighbor Lists} \label{sec:compmet_bah_neighlists}

Efficient distance calculation is a common problem in molecular dynamics simulations, since most force fields depend on interatomic distances in one way or another. A simple and common approach is to utilize cell lists~\cite{frenkel_daan_understanding_2002}, where the system is divided into smaller cubic cells, and atoms are assigned to these cells according to their coordinates. If the size of the cells is chosen properly with respect to the cutoff radius of the potential, checking for neighbors becomes simple: for each atom only atoms within the same cell and the directly neighboring cells need to be considered. 

It is possible to envision taking this approach to further dimensions, where we would now create cells not in coordinate space but in the higher-dimensional ACSF space. Unfortunately, this simple approach in unfeasible as the computational costs increase rapidly with dimensionality: in a one-dimensional system we need to check the central bin plus two neighbor cells, in two dimensions it is the central cell plus eight cells organized in a square, and so on with the total number of cells to be checked scaling as $3^D$ with $D$ the dimensionality of the space. This is clearly unfeasible for an ACSF set whose dimensionality starts at 20 but can contain as many as 100 ACSFs per atomic environment, and even cases with many hundred functions have been reported~\cite{P4945}.

In conclusion, cell-based neighbor lists efficiently reduce the degrees of freedom of the problem by creating cells, which we essentially also use for the binning step in the BAH algorithm. However, it rapidly fails when used in higher dimensions, which we avoid in our BAH algorithm by only finding points in ACSF space that are in the same bin/cell, and by utilizing hash tables to perform this check very efficiently using only a one-dimensional property for the comparison.

\subsubsection{The Naive Approach} \label{sec:compmet_bah_naive}

The naive approach to comparing atomic environments is to compare ACSF vectors for each pair of atoms directly. The only obvious simplification is that only atoms of the same element need to be compared. The performance of this procedure is very poor, since it scales linearly with the number of ACSFs, and quadratically with the number of environments in the data set, as for environment number $N$, we need to compare it with all the previous $N-1$ environments already processed.

Hashing and using hash tables solves this scaling problem, since lookup in a hash table is -- in principle -- a constant time operation~\cite{cormen_introduction_2009} that does not depend on the amount of data already stored in the table. Binning is needed before reaching this point, since similar floating point numbers would have very different hash values without a preparatory discretization step. 

\subsubsection{Binning} \label{sec:compmet_bah_binning}

Consequently, binning is the first step in the algorithm. The maximum and minimum values of each ACSF depend on the available data set and are known beforehand. For each ACSF, the resulting range is divided into an arbitrary number of intervals and the binning is done according to

\begin{equation}
    \label{eq:binning}
    B_j=\textrm{nint}\bigg(\textrm{divs}*\frac{\textrm{ACSF}_{\textrm{max}}-\textrm{ACSF}_{\textrm{val}}}{\textrm{ACSF}_{\textrm{max}}-\textrm{ACSF}_{\textrm{min}}}\bigg)
\end{equation}
where $B_j$ is the bin value for the $j$-th ACSF, nint is the nearest integer function, i.e., a round-off to the closest integer; and  $\textrm{ACSF}_{\textrm{max}}$, $\textrm{ACSF}_{\textrm{min}}$, and $\textrm{ACSF}_{\textrm{val}}$ are the maximum, minimum, and current value of the ACSF under consideration, respectively. The number of intervals is kept the same for all the ACSF types, although some of them might have larger or smaller ranges (see for example Fig.~\ref{fig:sf-histo}). A possible improvement to the binning procedure would thus be to aim for a certain density of ACSF values in each division, by tailoring the length and number of divisions to each ACSF.

This binning achieves multiple goals. In the first place, it transforms floating point numbers, which are imprecise and hard to hash, into integers. Floats should not be hashed directly because small changes in the accuracy of the floating point number representation, such as the limited precision when reading it from a file or small deviations resulting from rounding errors, give rise to very different hash values. Integers, on the other hand are easy to convert to a hash. 

Additionally, binning provides a sense of ``distance'' in the data set. Calculating distances directly from the difference between ACSF vectors suffers from the same scaling problems as the naive approach, and the usefulness of an Euclidean distance decreases with the size of the vector, as it becomes less unique and loses meaning~\cite{aggarwal_surprising_2001} as dimensionality increases. As the bins get smaller, fewer ACSF vectors will coincide, making the algorithm more sensitive only leaving those environments that are more and more similar in the same bucket.


Binning on its own does not solve the problem of the naive approach, since we would still need to do an all-against-all comparison of the individual bin vectors, with integers instead of floats. To solve this, a hash table is required, as described in the following section.

\subsubsection{Hashing and Hash Tables} \label{sec:compmet_bah_hashing}

Hash functions~\cite{cormen_introduction_2009} are a family of functions that can map data of arbitrary size to data of fixed size. In effect, a hash is a one-way function, that can assign an integer to any data type. This assignment is not unique as two objects that are different can result in the same hash value, i.e. a hash collision. This conversion is usually non-reversible such that if the hash is known, it is not possible to reconstruct the original object unless by brute force trial and error and comparing the resulting hashes. If two objects share the same hash (a ``hash collision''), they will usually be either exactly equal or very different, which is a desired property in some applications. Small changes to the input object will result in very different hash values, so the hash value in principle cannot be used directly as a measure of distance in input space. Hash functions are used in a variety of fields, such as in cryptography, where passwords as usually stored pre-hashed instead of in plaintext; or in the realm of data-validation and proofing such as in checksums, credit card numbers, bank routing numbers, ISBN book numbers, or blockchains. Hash functions make heavy use of the modulo function and byte-shifting operations.

The properties of a hash function allow us to create a hash table. A hash table resembles an array, but instead of assigning positions sequentially as in a normal array, positions to the hash table's ``buckets'' are assigned using the hash function. In effect, the hash function is used to index the hash table array using
\begin{equation}
\label{eq:hash_table}
    \textrm{index}=\textrm{hash} \% \textrm{array\_size} \quad ,
\end{equation}
where ``index'' is the index to be used when accessing the hash table array, ``hash'' is the hash function value of the object of interest, ``array\_size'' is the size of the array holding the hash table, and $\%$ is the modulo operator. The hash will always index an array position, no matter the size of the array.

One apparent problem arises here: The number of bins can reach up to $10^7$ subdivisions per ACSF. For the usual dozens to hundreds of symmetry functions required for a HDNNP data set, this amounts to a large amount of possible bin vectors that grows in a combinatorial fashion. How then is it possible to map all the possible bin vectors into a hash table of restricted size? As mentioned above, hash functions map larger spaces into smaller ones, so collisions are unavoidable. Various solutions exist for solving this problem~\cite{cormen_introduction_2009}, which are implementation dependent. One possibility, known as separate chaining, is to store all the collided keys in the same bucket as a list. Assignment to the hash table then consists of rapidly finding the correct bucket as in Eq.~\ref{eq:hash_table}, followed by a slower (but short) search through the list of key in this bucket. Another possibility, known as open addressing, is to assign keys to the first open bucket address if the current one is already occupied. Assignment of a new key then consists of using Eq.~\ref{eq:hash_table} to find an initial bucket (a fast operation), and then continuing through the bucket addresses until an unoccupied address is found (slower but a short process). Whatever the implementation utilized for collision resolution, it inflicts a computation overhead to all hash table operations, but if the number of collisions is kept low, this is not a problem. In normal operation every possible single bin vector will not be encountered since the data utilized to construct a HDNNP is not completely random, so this is not expected to involve much overhead.

An interesting feature of hashes is that this \textit{ansatz} results in a constant (when the number of hash collisions is not too high) search, assignment and insertion time of data into the table. In a normal array, if we want to check whether a new object is already present in the array, we need to traverse the array and compare element by element until it is either found and we stop the search early, or we reach the end of the array. In a hash table, we instead calculate the hash of the object and immediately check the corresponding position in the table.

This efficiency comes at the cost of some overhead: requiring more memory for storing the hash table since many buckets might be empty if the hash table is constructed with sequential memory positions, the need to precompute the hash for objects going into the table although hash calculations are usually fast, and dealing with hash collisions when they happen if we want to maintain unique buckets. Due to their properties, hash tables are a basic data structure in computer science~\cite{cormen_introduction_2009}, often utilized for efficient storage and retrieval of data.

A final advantage of hash tables for the use in this work is that they can be easily stored into a text file for future use. This way, a data set can be preprocessed into a hash table, and future structures can easily be compared against this record to detect repeated configurations. To store the hash table all that is needed is to write the unique binned integer vectors to the file (in an arbitrary order), optionally with a numeric ID associated to the structures in the data set that fall into that bucket of the table for an easy identification. To reconstruct the table, these binned arrays are read and used as members of a new table.

\subsection{Scaling} \label{sec:compmet_bah_scaling}

\begin{table}[]
    \centering
    \begin{tabular}{|c|c|}
    \hline
    Algorithm         & Scaling \\
    \hline \hline
    Naive Comparison  & $O(M*N^2)$ \\
    Binning           & $O(M*N)$\\
    Hashing           & $O(M*N)$\\
    Hash Table Lookup & $O(N)$\\
    \hline
    \end{tabular}
    \caption{Big $O$ notation scaling of the different algorithms under consideration. $N$ is the number of atoms corresponding to the number of atomic environments in the data set. $M$ is the number of functions in the atom-centered symmetry function vector.}
    \label{tab:on_scaling}
\end{table}

Next, we look at the scaling of the different parts of the algorithm in the big $O$ notation~\cite{cormen_introduction_2009}. This is important to realize why the naive approach soon becomes unfeasible and how the BAH algorithm improves on it. The results are summarized in Table~\ref{tab:on_scaling}. We will consider the case of searching once through a complete data set, and attempting to find repeated atomic environments.

In the following discussion, $N$ is the number of environments in the data set, i.e., the total number of atoms in all structures. $M$ is the number of functions in each ACSF vector corresponding to the dimensionality of our problem. We note that atoms of the same element always have the same ACSF sets, but this is not necessarily true for different elements. The scaling with respect to $N$ can be more important than regarding $M$, since the number of ACSF in a HDNNP is usually less than 100 per element for most systems, while the number of atomic environments can reach millions and has no upper bound.

The following scaling is observed:

\begin{itemize}
    \item Naive comparison and lookup: Comparison scales at worst as $O(M)$, since we need to compare each element in one ACSF vector to the corresponding element in another ACSF vector, but we might end early if a mismatch is detected. We then need to compare environment 1 with the next $N-1$ environments, environment 2 with the next $N-2$ environments and so on until environment $N-1$ for the last single comparison with environment $N$. This is a mathematical series that in the end scales as $O(N^2)$. Both parts of the algorithm together scale as $O(M*N^2)$.

    \item Binning: Binning scales with both the number of elements in each ACSF vector -- since we need to bin each element individually -- as $O(M)$. Additionally, it has to be done for each of the $N$ atomic environments ($O(N)$). Combined it scales as $O(N*M)$. This operation is usually very fast.

    \item Hashing: Hashing scales weakly with the size of the object being hashed ($O(M)$). There is some dependence on the specific implementation of the hash function (see Sec.~\ref{sec:results_bah_timings}) and the hashing needs to be repeated for each ACSF to be compared ($O(N)$). It is a comparably slow operation compared with a straight division in binning.

    \item Hash tables: Addition of data to a hash table and lookup are constant with respect to the size of the stored data set (which would be proportional to N), except for hash collisions $O(1)$. This is where the main time saving comes from. We have to repeat this $N$ times, once per hashed array, resulting in a scaling of $O(N)$.
\end{itemize}

Now we can estimate the total processing times. The naive case is simple, we need to perform $M*N^2$ operations to process the whole data set. For the BAH algorithm, we need to first bin the whole data set, then hash the resulting binned arrays, and finally store the result in the hash table detecting a collision if present. All of these times are additive since they are independent sequential operations. Putting this all together, we obtain 
\begin{equation}
    \begin{split}
    t_{\textrm{naive}} &= k_{\textrm{comp and lookup}}*O(M*N^2) \\
    t_{\textrm{bah}}   &= k_{\textrm{binning}}*O(M*N) + \\
                       &+ k_{\textrm{hashing}}*O(M*N) + k_{\textrm{hash lookup}}*O(N)
    \end{split} \quad ,
    \label{eq:on_timings}
\end{equation}
where each $k$ is the timing constant to perform that operation once, which depends on the actual implementation of each algorithm, the programming language of choice, and the CPU architecture. Notice that the naive approach shows the worst scaling, since it scales as $N^2$, with typical values of $N$ in the order of $10^4-10^6$. The BAH algorithm, on the other hand, consists of three linearly scaling additive components. This is tested in Section~\ref{sec:results_bah_timings} for an illustrative example, and the different timing constants estimated, for a Python implementation.

\subsection{Implementation} \label{sec:compmet_bah_implementation}

The algorithm has been implemented in Python 3.5, using the dict~\cite{python_data_structures} data structure, which is a hash table with the possibility to associate arbitrary data to each hash bucket. The set~\cite{python_data_structures} data structure is similar and can also be used, but can only store the hashed object and no other associated data. It can also be implemented easily in many other languages, since hash tables are a widely used data structure, and only pointers or allocatable arrays are needed to implement them from scratch. The dict object in Python already incorporates the step of hashing the data, so no explicit hash function is required in this case, and the actual implementation of the hash function is not relevant to the result as long as it avoids as many spurious collisions as possible.

The algorithms is straightforward to parallelize if this is required for larger data sets, or for non-synchronous processing, e.g. using a compute cluster associated with a database. This is due to the fact that hash tables can be easily combined. A central master process can hold the copy of the hash table, and dispatch binning and hashing operations to the slave processes; or each slave process can hold its own hash table and report back to a central process, which combines the slave sub-tables into a master hash table.

\section{Results} \label{sec:results_bah}

\subsection{Performance and Timings} \label{sec:results_bah_timings}

\begin{table}[]
    \centering
    \begin{tabular}{|c|c|c|}
        \hline
        \multicolumn{3}{|c|}{Naive} \\
        \hline
        Constant & Value (s/op$^2$) & op$^2$/s \\
        \hline
        $k_{\textrm{comp and lookup}}$ & 8.8E-8 & 11.000.000\\
        \hline
        \multicolumn{3}{|c|}{BAH} \\
        \hline
        Constant & Value (s/op) & op/s \\
        \hline
        $k_{\textrm{binning}}$ & 3.0E-6 & 336.000 \\
        $k_{\textrm{hashing}}$ & 1.8E-7 & 5.500.000 \\
        $k_{\textrm{hash lookup}}$ & 2.9E-7 & 3.400.000 \\
        $k_{\textrm{BAH global}}$ & 4.2E-6 & 238.000 \\
        \hline
    \end{tabular}
    \caption{Estimated scaling constants for the different parts of the naive and BAH algorithms, at a constant $M=10$ (scaling is assumed linear for other $M$ values, in the cases where relevant). Units are in seconds required per operation (s/op). The inverse constant is also given providing the number of operations per second (op/s). Note that the naive algorithm only seems ``faster'' because it is expressed in terms of op$^2$.}
    \label{tab:constants}
\end{table}

\begin{figure*}
    \centering
    \includegraphics[width=\linewidth]{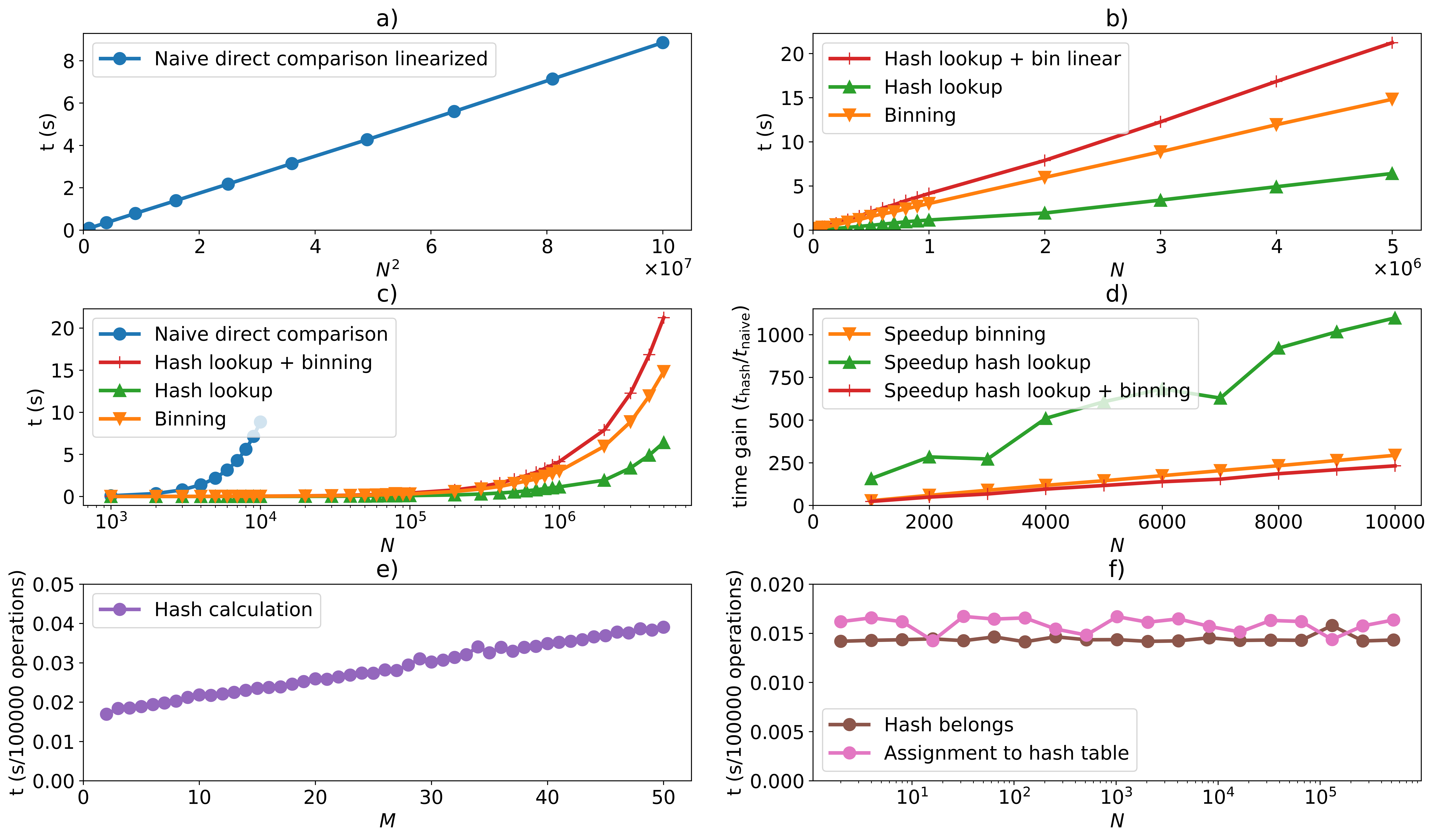}
    \caption{Plots of the timing of the different algorithms with increasing system size. a) Naive lookup vs. squared size of data set. b) Different parts of the BAH algorithm vs. size of data set. c) All algorithms together in log scale for comparison. d) Relative speedup or time gain of the different parts of the BAH algorithms compared to the naive approach, calculated as $t_{\rm algo}/t_{\rm naive}$, with $t_{\rm algo}$ the timings of the different parts of the algorithm from b). e) Scaling of the hash calculation with ACSF vector size, per 100000 operations. f) Behavior of hash table operations with data set size, per 100000 operations.}
    \label{fig:plotscaling}
\end{figure*}

For illustrative purposes, we present the timings and scalings of the naive and BAH algorithms on randomly generated values, as obtained from Python3.5 on a Intel Core i5-5300U CPU 2.30GHz. Fig.~\ref{fig:plotscaling} plots the behavior of the different algorithms for increasing data sets. 

As can be seen in Fig.~\ref{fig:plotscaling}a, the naive algorithm for the comparison of the atomic environments scales with the square of the data set size, while the BAH algorithm in Fig.~\ref{fig:plotscaling}b scales linearly. In the logarithmic scale of Fig.~\ref{fig:plotscaling}c combining the data of panels a) and b), it can be clearly seen that the costs of the naive algorithm increase much faster than those of the BAH algorithm. Fig.~\ref{fig:plotscaling}d shows the speedup (the relative time gain, $t_{\rm algo}/t_{\rm naive}$ for any of the sub-algorithms involved in BAH) between the BAH and the naive algorithms. Notice that this speedup \textit{increases} as the data set size increases, since the naive approach scales as the square of the data set size but the BAH scales linearly. Consequently, the larger the data set becomes, the faster the BAH approach becomes with respect to the naive approach. Fig.~\ref{fig:plotscaling}e shows that the hashing algorithms scales linearly with the size of the ACSF vector under consideration, but is extremely fast for typical vector dimensionalities. Finally, Fig.~\ref{fig:plotscaling}f confirms that, as expected, operations regarding the hash table object -- assignment to the hash table, and looking up if an object belongs to the hash table -- remain constant in time with data set size.

From these analyses and data we can estimate the different proportionality constants of  Eq.~\ref{eq:on_timings}, they are compiled in Table ~\ref{tab:constants}. Notice that the Naive and BAH halves of the table have different units. The fastest part of the BAH algorithm is the hash calculation ($k_{\rm hashing}$), while the bottleneck in the current implementation seems to be the binning ($k_{\rm binning}$). This is probably due to the division and rounding nearest integer operations involved in binning, and it could probably be improved with some vectorization or better numerical libraries. Not considered here is the required I/O to read ACSF data from a file, which might become a more serious bottleneck for larger data sets, but is however common to both algorithms. The values obtained here represent only an approximate order of magnitude since this will change significantly for different implementations and computing architectures.

\subsection{Analysis of the Distance in Symmetry Function Space}
\label{sec:results_bah_distance}

\begin{figure*}
    \centering
    \includegraphics[width=0.95\linewidth]{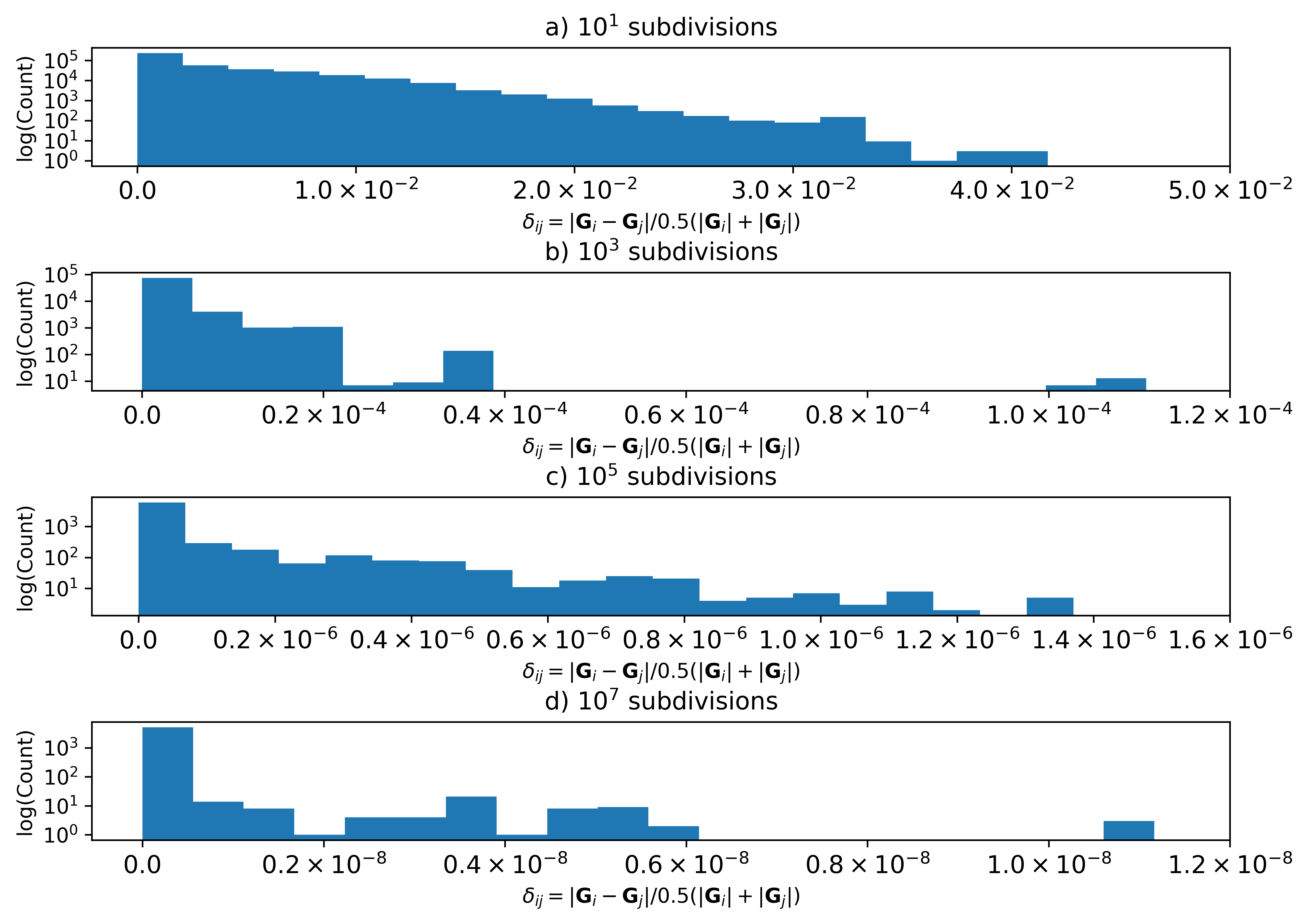}
    \caption{a)-d) Histograms for the typical intra-bucket ACSF relative distance ($\relacsfdist$) values for different subdivisions ($10^1, 10^3, 10^5, 10^7$) in the ZnO slab data set. Other intermediate subdivisions ($10^2, 10^4, 10^6$) exhibit similar behaviors. The counts axis is logarithmic for better visualization.}
    \label{fig:bah_gdist_histogram}
\end{figure*}

\begin{figure}
    \centering
    \includegraphics[width=0.95\linewidth]{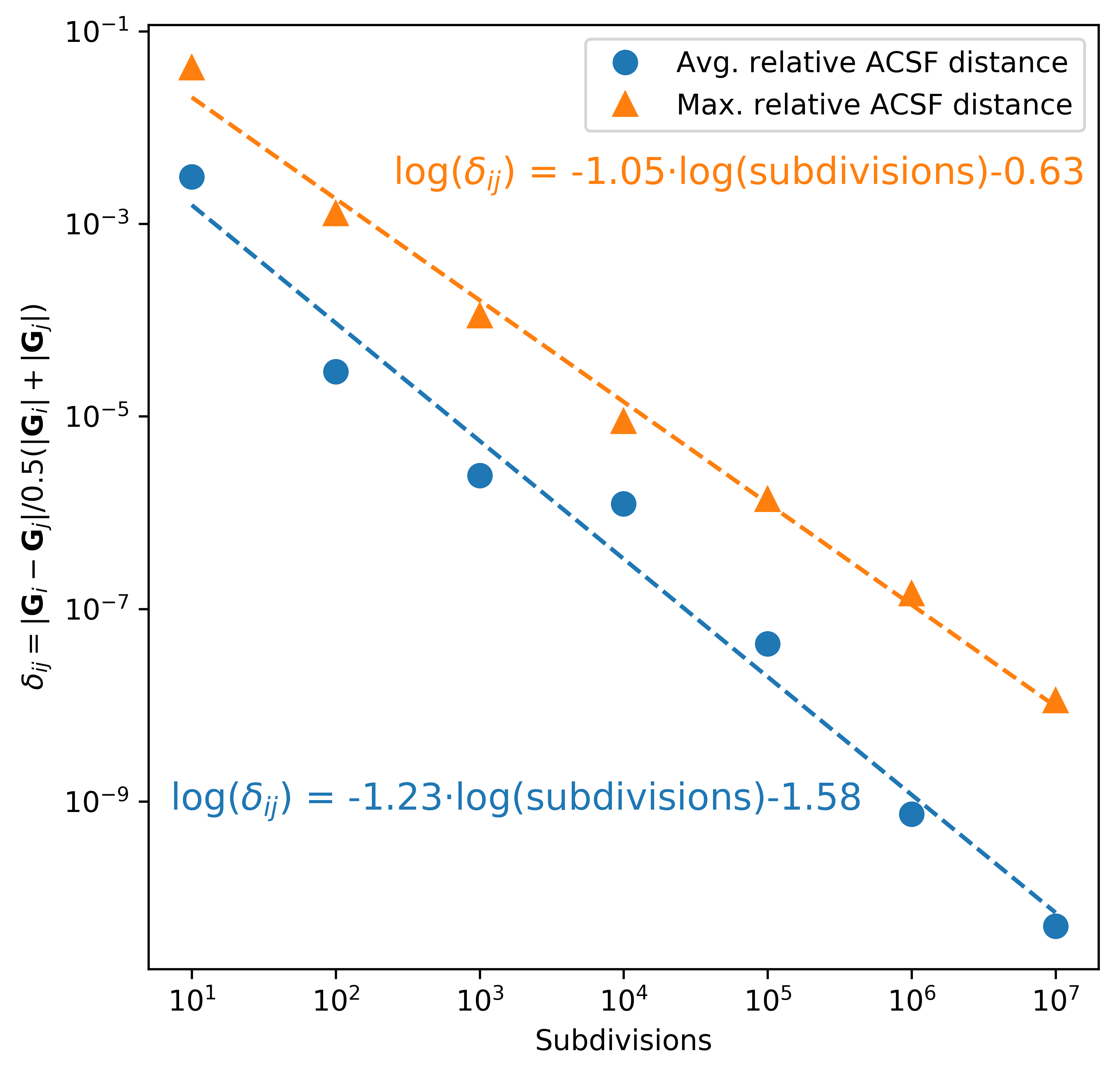}
    \caption{Maximum and average intra-bucket relative distances for the histograms in Fig.~\ref{fig:bah_gdist_histogram} versus number of subdivisions, in log scales. Notice that they follow  approximately linear relationships, and trendlines with corresponding fitting equations are included.}
    \label{fig:bah_gdist_trendline}
\end{figure}

\begin{figure*}
    \centering
    \includegraphics[width=0.95\linewidth]{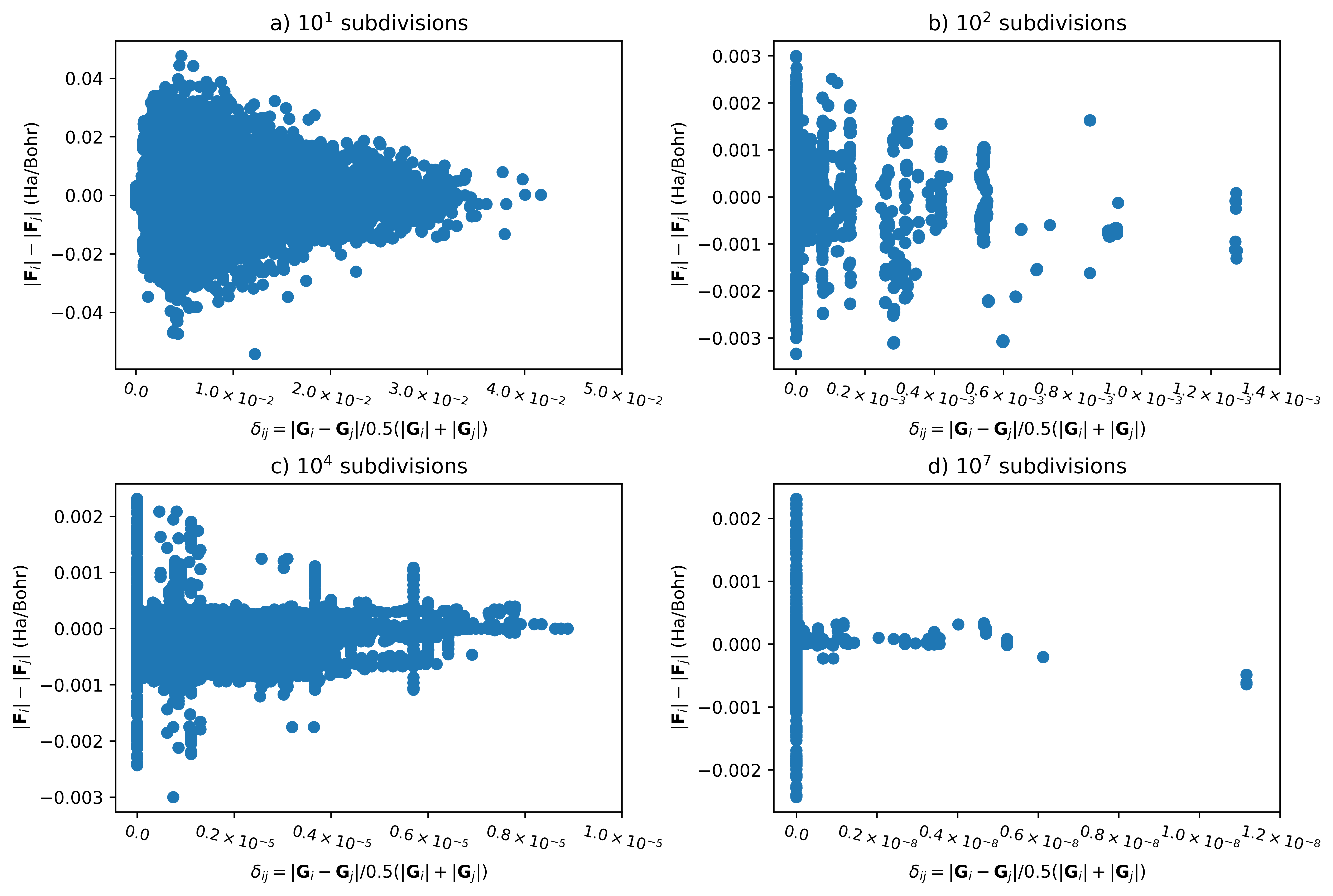}
    \caption{Difference in force magnitude vs. the ACSF relative distance, $\relacsfdist$, for different subdivisions of the BAH algorithm applied to the ZnO slab data set. The points present in each subplot are not always the same, since the plots are generated from environments that collided for a given number of subdivisions. Notice the difference in the scale of the X and particularly the Y-axis for a) when compared to b)-d); the force spread for structures with $\delta_{ij} \approx 0$ is due to remaining numerical noise in the DFT data.}
    \label{fig:fdistance_vs_gdistance}
\end{figure*}

An interesting question is how the algorithm reflects distances in ACSF space, since some information is lost in the process of binning and hashing of the atomic environment vectors. Hashes themselves are not a useful measure of distance since the resulting hash is not smoothly continuous with respect its inputs, but we would expect similar ACSF vectors to end in the same bucket. A reliable binning of only similar structures is an important condition for the BAH method to be useful.
For this purpose, we now investigate all the ACSF vector distances obtained for atomic environments that fall in the same bucket using different subdivisions of the ACSF space. We define a relative distance in ACSF space, $\relacsfdist_{ij}$ between atoms $i$ and $j$ of the same element, as 
\begin{equation}
    \relacsfdist_{ij} = \frac{|\mathbf{G}_i-\mathbf{G}_j|}{0.5(|\mathbf{G}_i|+|\mathbf{G}_j|)}
\end{equation}
where $\mathbf{G}_i$ and $\mathbf{G}_j$ are a pair of symmetry function vectors corresponding to atomic environments that ended up in the same bucket, and which are thus similar for the BAH algorithm. We plot a histogram of the calculated distances in Fig.~\ref{fig:bah_gdist_histogram} for different subdivision numbers. Most of the distances in the histogram are close to zero as expected. Notice that as we increase the number of subdivisions, the maximum intra-bucket distance drops quickly due to the more stringent criterion for structural similarity in the binning process, becoming close to the floating point noise (either due to the limited precision of floating point numbers in a computer representation a.k.a. the ``machine epsilon'', or the limited precision of data such as coordinates and ACSF values held in text format) for the maximum number of subdivisions such that the differences for many subdivisions are probably due to round-off errors and float-to-string conversions rather than significant distances in ACSF space. Consequently, the histograms show that the BAH algorithm is indeed closely correlated to distances in ACSF space, up to a given maximum distance depending on how the multi-dimensional space is subdivided for the binning step.

Interestingly, as shown in Fig.~\ref{fig:bah_gdist_trendline}, the maximum and average $\relacsfdist$ obtained from these histograms follow a linear relationship with the number of subdivisions, on a double logarithmic scale. Therefore, changing the subdivisions parameter allows us to fine-tune the maximum detected atomic environment distance in a predictable way.

Given this behavior of the distances in ACSF space, it is also of interest to study the corresponding behavior of the properties associated to each atomic environment such as the atomic forces. In Fig.~\ref{fig:fdistance_vs_gdistance} we plot the difference in force magnitude~\footnotemark[1] vs. the ACSF relative distance, $\relacsfdist$, for different subdivisions. As shown in a), there is a relationship between the two quantities, since one would expect that atoms whose environments/ACSF vectors are similar should also present similar forces. Despite this, the relationship is not strong, since distances in ``force space'' do not necessarily transfer linearly into ACSF space~\cite{parsaeifard_assessment_2020}. As the number of divisions increases and the force vectors considered correspond to closer environments, the force distance quickly falls. In the end (d), this force distance corresponds to the numerical noise present in the reference DFT data, since he environments detected are actually identical (up to numerical noise).

\subsection{Results for Different Divisions and Symmetry Functions}
\label{sec:results_bah_sfsets}

\begin{figure*}
    \centering
    \includegraphics[width=\linewidth]{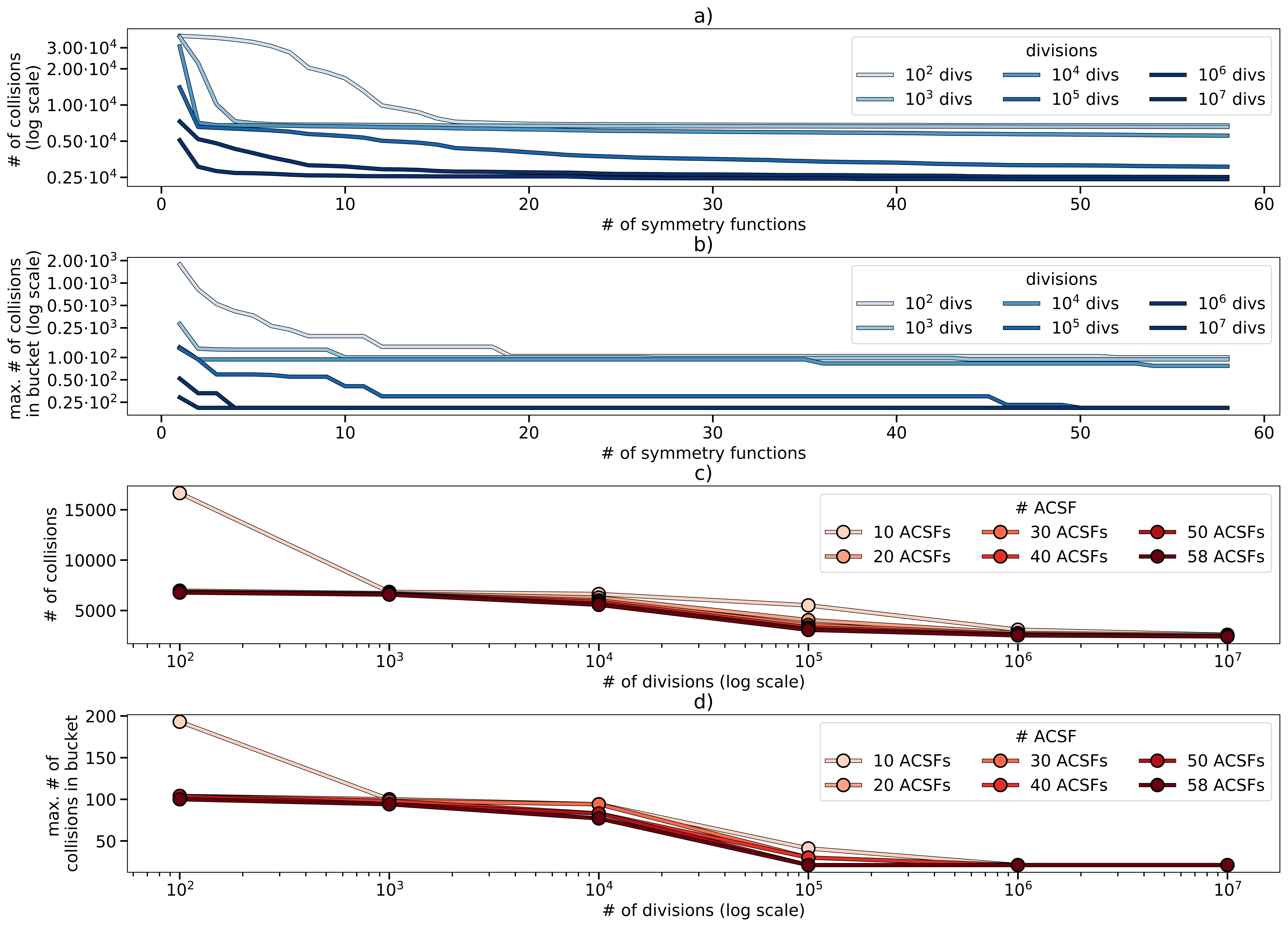}
    \caption{ Panels a) and b) show the total and maximum number of hash table collisions, i.e., configurations that hash into the same bucket due to similarity of their ACSF vectors, vs. number of ACSFs, for different binning divisions. Panels c) and d) show the same properties as a function of the number of binning divisions, for different numbers of ACSFs.}
    \label{fig:bahstats}
\end{figure*}

An interesting question is how the resolution power of the algorithm, i.e., the ability to differentiate  ACSF vectors, changes as we increase the number of binning subdivisions, and as we change the ACSF descriptor set itself. For this purpose, we have analyzed the ZnO (10$\bar{1}$0) slab data set.

A count of collisions was performed on this data set, which as described before occur when two environments end up in the same hash table bucket, due to their binned vectors being the same, which implies their original ACSF vectors were at least similar. We keep track of the total number of collisions, and the maximum number of collisions in a single bin, for different divisions and an increasing ACSF set.

We would expect both total and maximum number of collisions to go down as both divisions and numbers of ACSFs increase, since more divisions means that environments need to be more similar in ACSF space to collide (see Sec.~\ref{sec:results_bah_distance}) and more ACSFs lead to a more granular description of each environment. Eventually, this count converges as we are left with only the environments that are exactly the same, which can happen in a data set due to repeated parts of a configuration for example, if parts of a slab far away from a chemically modified region remain essentially constant. This is in fact found in Fig.~\ref{fig:bahstats}. Here we have performed the BAH analysis on an increasing number of ACSFs, in the order presented in the supporting information.  

In this figure we note that in a), collisions go down extremely quickly as we increase the ACSF descriptor set, and then plateau with a slight downward trend that is hard to observe due to the scale of the plot. The line with $10^5$ divisions seems to offer the most granularity, showing changes across the whole ACSF set under consideration. Being able to differentiate chemical environments is a necessary (but not sufficient) condition for a good HDNNP fit, in which case the BAH algorithm could be utilized to identify a minimum floor to the size of the ACSF set.

At this point, the question arises of which subdivision range is ``best'' to describe a given data set, and whether this is actually dependent on the specific data set. As can be seen from Fig.~\ref{fig:bah_gdist_trendline}, the number of subdivisions roughly corresponds to the symmetry function space distance between the collided atomic environments. As such the ``right'' subdivision range depends on whether we want to detect environments that are only roughly similar or exactly the same, and there is not a single ideal value. For the type of analysis presented in Fig.~\ref{fig:bahstats}, a lower number of subdivisions (in the range of $10^2$ to $10^4$) provides a more granular behavior in the number of collisions vs. symmetry functions utilized, which results in an easier to analyze trend. For detecting contradictions (see Sec.~\ref{sec:results_bah_contradictions} we require environments that are either extremely similar or exactly the same, in which case the upper range of subdivisions ($10^6$ to $10^7$) is better suited. 

Whether the number of subdivisions required depends on the specific data set is harder to evaluate. Since our data sets are derived from physically ``reasonable'' configurations corresponding to chemical systems, they share roughly the same properties, with some differences depending on the involved elements, states of matter present, energy ranges covered, etc. The parameters of the trendlines in Fig.~\ref{fig:bah_gdist_trendline} might depend on the specific composition of the data in the data set, but as long as the relationship with ACSF space distance remains, the specific parameters are not crucial.

In the end no specific number of subdivisions is ideal for every situation, and this has to be tested with each data set and adapted to each desired analysis, but the BAH process is so fast that binning a data set multiple times is not a problem. Our recommendation is to test three widely separated orders of magnitude of subdivisions ($10^3-10^5-10^7$), and refine according to the results.

\subsection{Comparison of Atomic Environments and Conflicting Information}
\label{sec:results_bah_contradictions}

\begin{figure*}[h!]
    \centering
    \includegraphics[width=\linewidth]{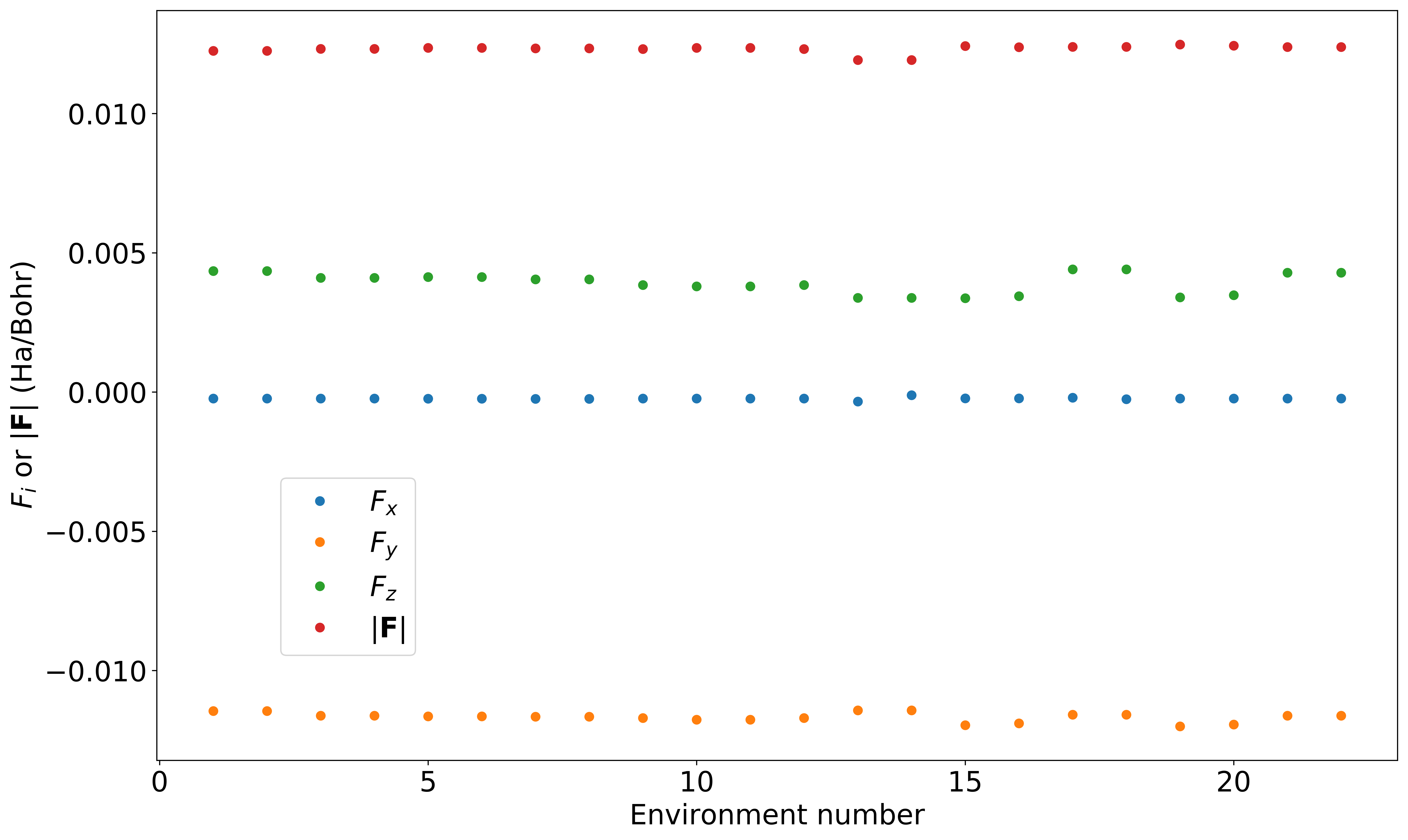}
    \caption{Force components and force vector magnitude for 22 environments found in a collision bucket. Note that although the ACSF vector for all environments is identical, there are slight differences in force values arising from numerical noise in the DFT calculations.}
    \label{fig:contradictions}
\end{figure*}

The result of running the BAH algorithm is a list of environments that fall into the same bucket. That is, we obtain a list of collisions representing structurally similar atomic environments as defined above. This is valuable information and can be used to predict if a new configuration obtained from a simulation employing the HDNNP is sufficiently different from the available data to justify an inclusion in the reference data set to refine the potential. All the atomic environments in a large number of structures structure obtained in long validation simulations can be screened in this way, and for a most efficient use of subsequent electronic structure calculations it is possible to identify those structures from this pool, in which the highest fraction of environments is sufficiently different for the existing reference data.

Another possibility is the search for contradictions in the data set. Contradictions in this case means atoms whose ACSF sets are similar, but their derived properties (any per atom predicted property, such as force, spin, charge, etc.) differ by more than an acceptable threshold. This could be due to a too small ACSF set or cutoff radius of the ACSFs, which does not allow to correctly distinguish chemically different atomic environments, due to the neglect of long-range interactions beyond the cutoff radius, or due to incorrect electronic structure data resulting, e.g., from a poor convergence level. Contradictions are detrimental to the fitting process, since in case of conflicting data the HDNNP cannot reach a high fitting accuracy~\cite{behler_atom-centered_2011}.

If we apply this analysis to our data set, with $10^5$ binning divisions we find that the bucket with most collisions contains 22 environments. The ACSF vector of these configurations is identical, but plotting their DFT force components~\footnotemark[1] and magnitude results in Fig.~\ref{fig:contradictions}. We can see that the forces are not exactly identical, but they are within the expected error margin for the HDNNP~\cite{weinreich_properties_2020}, i.e. below about 100 meV/Bohr. In this case, no contradiction is detected, but in other situations we found structures that have not properly been converged for various reasons. Identifying and eliminating these data substantially improved the HDNNPs in this case. For larger data sets, the points within buckets could be automatically analyzed, and a contradiction warning raised if the force difference is above a given threshold.

\footnotetext[1]{Note: When comparing force components directly, care should be taken. ACSF vectors are invariant with respect to rotations and translations in coordinate space, but forces are \textit{not}. This is due to the derivatives involved in going from energy to forces, which add a direction component. The result is that with the same ACSF vector, one can have different force vector orientations, that is, the components of the force vector might not match. The predicted magnitude of the force vector should on the other hand remain consistent since it is directionless. A trivial example of this is an unrelaxed unmodified slab with two interfaces: atoms in the top and bottom surfaces will have identical environments as described by their ACSFs, but the $Z$-component of their force vectors will necessarily, due to symmetry, be opposite. This becomes more complicated for more homogeneous systems such as liquids and amorphous solids, where the same atomic environment might be found in a variety of orientations. Thus only force vector magnitudes should be compared, or a consistent orientation of the environments should be achieved in some way.}

\section{Supporting Information} \label{sec:supporting_information}

In the supporting information we present:
\begin{itemize}
    \item A list of ACSF parameters for the studied ZnO slab data set.
    \item The code utilized to perform the scaling tests in Sec.~\ref{sec:results_bah_timings}. 
\end{itemize}

\section{Conclusions} \label{sec:conclusion}

In this work we have presented a bin and hash method, which allows a computationally very efficient comparison of a large number of geometric atomic environments, which are used in the construction of modern machine learning potentials. In case of high-dimensional neural network potentials, which we use as a typical example here, these environments are usually described by vectors of atom-centered symmetry functions. We show that the ability of the method to identify similar atomic environments can be systematically controlled by the number of subdivisions used in the binning process of the ACSF vectors, but also a large number of alternative descriptors proposed in the literature is equally applicable. 

The method is fast, simple and robust with many applications in the construction of machine learning potentials. One example is the identification of redundant atomic environments in the reference data sets used for the construction of the potential as a basis for the decision which structures should be included in the training set. This is an essential step, as a systematic coverage of the configuration space is very important for obtaining reliable potentials, which an excessive amount of data would turn the construction and use of the potentials unfeasible. Due to the use of hash functions and tables, the method can process millions of candidate atomic environments in a number of minutes, being much faster than a naive direct comparison approach. The obtained information can be stored in data libraries that can be efficiently searched at a later stage if needed. We note that in this context the BAH algorithm is complementary to the use of active learning, as the BAH algorithm is based on the geometric structure and its description, while it does not require the availability of trained ML potentials as no property evaluations are needed. Active learning on the other hand is based on the comparison of predicted properties, which allows to focus on the reliability of the target property, while it depends on the availability of preliminary models and their evaluation.  

Another application is the validation of the structural resolution capabilities of the descriptors used for the discrimination of different atomic environments. Poor descriptor sets result in a large number of environments appearing erroneously to be structurally similar although local physical properties like forces substantially differ.
Finally, the method can be used to identify conflicting data in the training set, which might result from an insufficient convergence level of the reference electronic structure calculations and other types of errors resulting in inconsistent information. 
Consequently, the bin and hash method has been found to be a useful tool for solving a variety of challenges emerging in the construction of machine learning potentials, with  many additional potential applications in other fields requiring the efficient comparison of structural features, such as genetic algorithms~\cite{deaven_molecular_1995}, minima hopping~\cite{goedecker_minima_2004}, and kinetic Monte Carlo~\cite{voter_introduction_2007} simulations.

\begin{acknowledgments}

We thank the Deutsche Forschungsgemeinschaft (DFG) for financial support (Be3264/10-1, project number 289217282 and INST186/1294-1 FUGG, project number 405832858). JB gratefully acknowledges a DFG Heisenberg professorship (Be3264/11-2, project number 329898176). We would also like to thank the North-German Supercomputing Alliance (HLRN) under project number NIC00046 for computing time. 
\end{acknowledgments}

\clearpage 
\bibliography{bibliography.bib}

\end{document}